\documentstyle[psfig]{mn}
\newcommand\etal{et al. }
\newcommand\refer{\par \noindent\hangindent=3pc \hangafter=1}

\newcommand\ros{{\it ROSAT}}

\newcommand\eg{eg }

\newcommand\ie{i.e. }

\title{The fraction of galaxies that contain active nuclei and their accretion rates}
\author[M.J. Page]{M.J. Page\(^{1}\)\\
\(^{1}\)Mullard Space Science Laboratory, University College London,
Holmbury St Mary, Dorking, Surrey RH5 6NT, UK.}

\date{}

\begin{document}
\maketitle

\begin{abstract}
 
We investigate the relationship between the present day optical
luminosity function of galaxies and the X--ray luminosity function of
Seyfert 1s to determine the fraction of galaxies which host Seyfert 1
nuclei and their Eddington ratios. 
The 
local type 1 AGN X--ray luminosity function is well
reproduced if \(\sim 1\%\) of all galaxies are type 1 Seyferts which
have Eddington ratios of \(\sim 10^{-3}\). However, in such a model
the X--ray luminosity function is completely dominated by AGN in E and
S0 galaxies, contrary to the observed mix of Seyfert host galaxies. 
To obtain a plausible mix of AGN host galaxy
morphologies requires that the most massive black holes in E and S0
galaxies accrete with lower Eddington ratios, or have a lower
incidence of Seyfert activity, than the central black holes of later
type galaxies.

\end{abstract}
\begin{keywords}
galaxies:active -- galaxies:luminosity function -- accretion
\end{keywords}
\section{Introduction}

Accretion onto a massive black hole has remained the standard paradigm
for active galactic nuclei (AGN) for several decades (\eg Lynden-Bell
1969, Rees 1984). In recent years, a considerable amount of evidence has
accumulated for the existence of massive black holes in many, perhaps
all, galaxies (Magorrian \etal 1998, Kormendy \& Richstone 1995).  The
connection between the formation and evolution of massive black holes
and their host galaxies is currently the subject of great interest and
investigation (\eg Fabian 1999, Salucci \etal 1999, Cattaneo, Haehnelt
and Rees 1999).

The luminosity function of AGN has been used for many years to track
the statistical evolution of the AGN population with cosmic epoch, but
does not allow distinction between a relatively small population of
long lived AGN or many short lived generations of AGN. The Eddington
ratio, \ie the ratio of an object's luminosity to its Eddington
luminosity, was proposed to be a powerful discriminator of AGN
activity patterns by Cavaliere \& Padovani (1988) with long lived AGN
having low Eddington ratios (\(\sim 10^{-4}\)) and short lived AGN
having high Eddington ratios (\(\sim 1\)).  In a variety of wavebands,
from radio to X--ray, the luminosity function of AGN has a
characteristic two-power law shape with a knee dividing the low and
high luminosity objects (\eg Dunlop \& Peacock 1992, Boyle, Shanks \&
Peterson 1988, Page \etal 1997). It has an obvious resemblance to the
luminosity function of galaxies, which is not surprising since AGN are
found in galaxies. Only a few attempts have been made to relate the 
galaxy and AGN luminosity functions, predominantly in the form of models 
for the joint formation of galaxies and AGN (\eg 
Kauffmann \& Haehnelt 2000, Monaco, Salucci, \& Danese 2000 and 
Haehnelt \& Rees 1993) yet the direct comparison of galaxy and AGN 
luminosity functions  can shed light on two
fundamental properties of the black hole population: the fraction of
massive black holes which are active, and their Eddington ratios.
This is the subject of this paper.

Throughout we have taken \(q_{0}=0\) and \(H_{0}=100\ h\ {\rm km\ 
s^{-1} Mpc^{-1}}\).

\section{Method}

\subsection{The relationship between galaxy and AGN luminosity functions}

The luminosity function of AGN must be related to
the mass function of massive black holes. If the masses of central
black holes are related to the masses of their host galaxy bulges (Magorrian
\etal 1998, Kormendy \& Richstone 1995), then the
luminosity functions of AGN and galaxies must be strongly related. This is
modeled in the following formalism.

We define the luminosity function as 
\[\phi=\frac{d^{2}N}{dVdL}\]
Where $N$ is number of objects, $V$ is comoving volume and $L$ is luminosity.
and start from the luminosity functions \(\phi_{i}\) of galaxies of \(n\)
different morphological types \(i\). We assume that the spheroid
components of galaxies of type \(i\) produce a fraction of their light
\(f_{sph}(i)\).  We can then translate from the \(\phi_{i}\)s to a
spheroid luminosity function \(\phi_{sph}\) by:
\[\phi_{sph}(L_{sph}) = \sum_{i=1}^{n}
\frac{\phi_{i}(L_{sph}/f_{sph}(i))}{f_{sph}(i)}\]
We then obtain the mass function of spheroids by:
\[
\frac{d^{2}N}{dVdM_{sph}} = \phi_{sph} \frac{dL_{sph}}
{dM_{sph}}
\]
and a black hole mass function \(\Phi_{BH}\) is obtained using
\[
\Phi_{BH} = \frac{d^{2}N}{dVdM_{BH}}=
\int \frac{d^{2}N}{dVdM_{sph}} P(M_{BH} \mid M_{sph}) dM_{sph}
\]
where \(P(M_{BH} \mid M_{sph})\) is the distribution of nuclear black
hole masses given spheroid mass \(M_{sph}\).

Some fraction \(f_{AGN}\) of nuclear black holes are in an active
(luminous) state and emit with luminosity \(L_{AGN}\).
Both \(L_{AGN}\) and \(f_{AGN}\) are likely to depend on \(M_{BH}\).
For simplicity we assume a functional relationship 
\(L_{AGN} = F(M_{BH})\), although a distribution of luminosities
\(P(L_{AGN} \mid M_{BH})\) would be more realistic.

The AGN luminosity function \(\phi_{AGN}\) is then:
\begin{equation}
\phi_{AGN}=
f_{AGN}
\Phi_{BH}
\frac{dM_{BH}}{dL_{AGN}}
\label{eq:phiagn}
\end{equation}

It is appropriate to formulate the ratio of \(L_{AGN}\) to \(M_{BH}\) in terms
of the Eddington ratio \(\epsilon\), 
the ratio of bolometric to Eddington luminosity
\(L_{\rm E}=1.3 \times 10^{38} (M_{BH}/M_{\sun}) {\rm erg s^{-1}}\). 
\begin{equation}
\epsilon=\frac{L_{AGN}}{M_{BH}}\ \frac{M_{\sun}}{1.3 \times 10^{38} 
{\rm erg\ s^{-1}}}
\label{eq:eddingtonratio}
\end{equation}

\section{Constructing the black hole mass function}
\label{sec:massfun}

For the local galaxy luminosity functions we have chosen to use the recent
determinations from the 2DF galaxy redshift survey (Folkes \etal 1999). This
provides Schechter function model luminosity functions for 5 different
spectroscopic classes of galaxy 
(corresponding to different morphological types)
 at \(z<0.2\). We have used values for \(f_{sph}\) (the fraction of B band light due to
the spheroid component) based on the values given by Meisels \& Ostriker
(1984).

The model luminosity functions and spheroid fractions are listed in Table
\ref{tab:galaxylfs}.

\begin{table}
\caption{Schechter model galaxy luminosity functions from Folkes \etal (1999)
and assumed fraction $f_{sph}$ of B band light due to the spheroidal
component.}
\label{tab:galaxylfs}
\begin{tabular}{lcccc}
Type&$M_{B}^{*}$&$\alpha$&$\phi^{*}$&$f_{sph}$\\
E/S0&-19.61&-0.74&9.0&0.7 \\
Sab &-19.68&-0.86&3.9&0.25\\
Sbc &-19.38&-0.99&5.3&0.15\\
Scd &-19.00&-1.21&6.5&0.1 \\
Sdm &-19.02&-1.73&2.1&0.02\\
\end{tabular}
\end{table}

For the spheroid mass - to light ratio we use the best fit relation of 
Magorrian
\etal (1998) for V:
\[
M/M_{\sun}=0.097\ h\ (L/L_{\sun})^{1.18}
\]
and assume B-V=0.9 for the spheroid component of all galaxy types (Pence 1976).
For the black hole to spheroid mass
distribution we use the best fit log-gaussian relation from Magorrian \etal
(1998):
\[
P(M_{BH} \mid M_{sph})=
\frac{f_{BH}\ e^{-0.5 [{\rm log}(M_{BH}/M_{S})-{\rm log}(x_{0})]^{2}/\Delta^{2}}}
{M_{BH} \Delta \sqrt{2 \pi}\ {\rm log_{e}}(10)}
\]
where \(\Delta = 0.51\), \(f_{BH}=0.97\) and \(x_{0}=5.2 \times
10^{-3}\). We truncate this distribution at \(\pm 3\sigma\)
because otherwise it predicts an unrealistically large population of
very massive (\(> 10^{11} M_{\sun}\)) black holes.

\begin{figure}
\begin{center}
\leavevmode
\psfig{figure=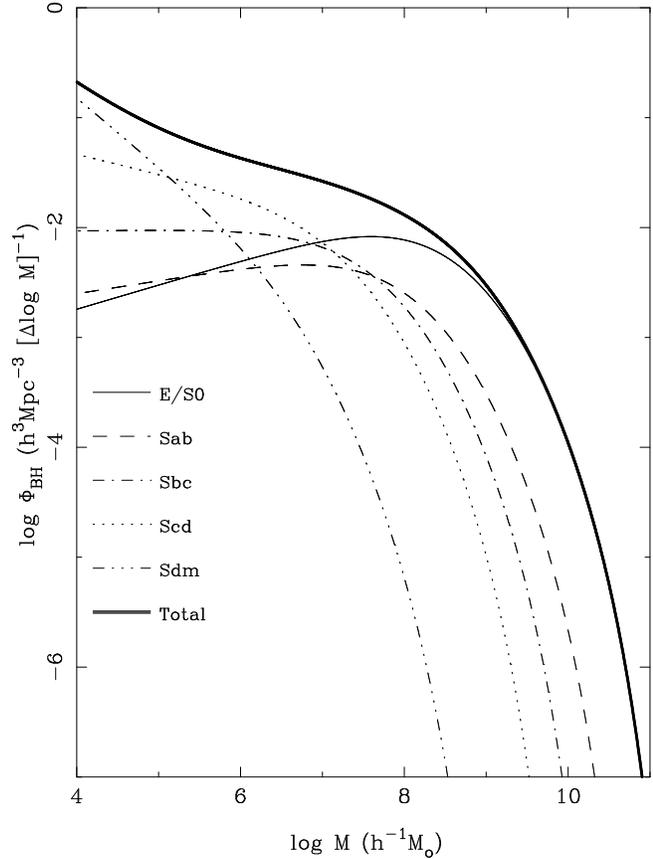,width=85truemm}
\caption{Mass function of black holes derived in Section \ref{sec:massfun}.}
\label{fig:massfun}
\end{center}
\end{figure}

The resultant black hole mass function is shown in Fig. \ref{fig:massfun}.

\section{Fitting the accretion rate and active fraction}
\label{sec:fitting}

For the AGN luminosity function we have chosen to use the extended
{\it Einstein} medium sensitivity survey (EMSS) X--ray selected sample
(Stocke \etal 1991) restricted to type 1 (broad line) Seyferts with
\(z < 0.2\).  X--ray selection is ideal because it is insensitive to
the optical host galaxy properties.  The restriction to broad line
objects ensures that the sample contains only unobscured AGN for which
the observed X--ray flux is a good measure of intrinsic luminosity,
and the redshift restriction ensures that the sample is well matched
to the 2DF galaxy sample and is without significant cosmological
evolution.

The simplest relation between black hole mass and AGN luminosity (and
therefore the starting point for this investigation) is obtained by
assuming an Eddington ratio \(\epsilon\) and an active fraction
\(f_{AGN}\) independent of mass or galaxy type; this will be referred to as Model A henceforth. Since we are constructing an
X--ray luminosity function rather than a bolometric luminosity function, and
only considering type 1 Seyferts rather than all AGN, we replace \(f_{AGN}\)
with \(f_{S1}\) and
reformulate Equation \ref{eq:phiagn} in terms of the Eddington ratio and 
include a bolometric correction \(\beta\), where \(\beta\) is the ratio of the
bolometric luminosity to the X--ray luminosity. 
\begin{equation}
L_{X}= \frac{0.013\epsilon M_{BH}}{\beta}
\end{equation}
where \(L_{X}\) is the AGN 0.3-3.5 keV luminosity in units of \(10^{40} {\rm
erg s^{-1}}\) and \(M_{BH}\) is in units of \(M_{\sun}\).
\begin{equation}
\phi_{X}=\frac{d^{2}N}{dVdL_{X}}=
\frac{f_{S1}\beta}{0.013\epsilon}
\frac{d^{2}N}{dVdM_{BH}}
\end{equation}
We adopt \(\beta=20\) based on the mean type 1 AGN spectral energy
distribution from Elvis \etal (1994). 

A binned (\(\Delta \log L = 0.3\)) AGN X--ray luminosity function was computed
from the EMSS sample (Fig. 2a) 
and compared to the model predicted luminosity function
using the method described in Sections 2.3 and 5.2 of Page \& Carrera
(2000). Only luminosity bins containing \(> 10\) objects were used in the
fitting so that \(\chi^{2}\) could be used as a goodness of fit estimator; the
resultant binned luminosity function uses 152 EMSS AGN.

\begin{figure*}
\begin{center}
\leavevmode
\psfig{figure=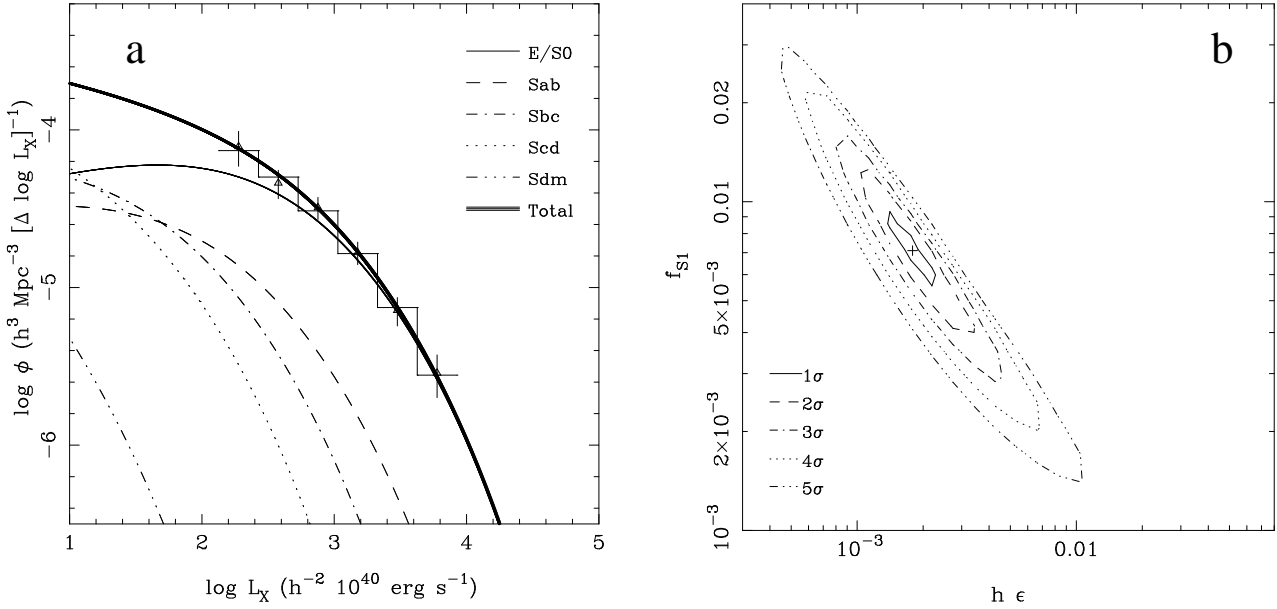,height=80truemm,angle=270}
\caption{(a) X--ray luminosity function of the EMSS $z<0.2$ AGN and best fit 
model assuming a universal Eddington ratio and a Seyfert 1 fraction
\(f_{S1}\) which is independent of galaxy type or black hole mass (Model A). 
(b) $\chi^{2}$ confidence contours for the values of Eddington ratio
$\epsilon$ and active Seyfert 1 fraction $f_{S1}$.}
\label{fig:lumifun}
\end{center}
\end{figure*}

A very good fit to the X--ray luminosity function of type 1
AGN is easily found. The best fit has \(h \epsilon=1.8\pm0.5 \times
10^{-3}\) and \(f_{S1}=7\pm2 \times 10^{-3}\) 
(where the errors define a box containing the \(1 \sigma \), 
\(\Delta \chi^{2} = 2.3\), confidence interval) with an extremely low
\(\chi^{2}\) of 0.4
for 6 fitted data points and 2 free parameters (\ie 4 degrees of freedom, so \(\chi^{2}/\nu=0.1\)). 
The best fit luminosity function is shown in Fig. \ref{fig:lumifun}a; the
predicted luminosity function passes almost exactly through the data. Fig. 
\ref{fig:lumifun}b shows \(\Delta \chi^{2}\) confidence contours on the
fitted parameters.

However, despite the very good fit, there is a problem with this
simple model.  It is seen in Fig. \ref{fig:lumifun}a that the EMSS
luminosity function is produced almost exclusively by E and S0
galaxies with only a small contribution from Sab galaxies at the low
luminosity end, but it is well known that Seyferts are often
spirals. Indeed, spirals are a significant component of the EMSS
Seyfert sample itself; this is demonstrated in
Table \ref{tab:EMSSmorphs} which gives the morphologies of the nearest
EMSS Seyferts.

\begin{table}
\caption{The nine closest galaxies from the EMSS Seyfert 1 sample, with their
de Vaucouleurs (1959) morphological classifications as listed in the 
NASA
Extragalactic Database (NED).}
\label{tab:EMSSmorphs}
\begin{tabular}{lcc}
Source & $log L_{X}$ & morphology\\
MS1215.9+3005 & 2.18 & SB(s)a    \\    
MS0339.8-2124 & 2.30 & SB(rs)a   \\    
MS0459.5+0327 & 2.59 & E         \\    
MS1158.6-0323 & 2.45 & E         \\    
MS2252.2+1126 & 2.00 & Sab       \\    
MS1846.5-7857 & 2.08 & SAB(r)b   \\    
MS1136.5+3413 & 2.74 & SB0	 \\    
MS0048.8+2907 & 2.96 & SB(s)b    \\    
\end{tabular}
\end{table}

To reproduce the the EMSS luminosity function with a mix of
morphological types requires additional complexity in the model:
to prevent early type galaxies completely dominating the luminosity function,
\(f_{S1}\) and/or \(\epsilon\) must depend on \(M_{BH}\) and/or
galaxy morphology. Since different galaxy morphologies dominate the
mass function at different masses, a direct dependency of \(f_{S1}\) or 
\(\epsilon\) on \(M_{BH}\)
results in an indirect dependency on morphology and vice-versa.  
If \(\epsilon\) is
to be varied it must be in such a way as to steepen the model
luminosity function so that the relative contribution of E/S0 galaxies
is reduced in the EMSS luminosity range. This means that more massive
black holes (in E/S0 galaxies) must accrete with lower Eddington
ratios than those of lower mass. If \(f_{S1}\) is to be varied it
must be such that a smaller proportion of more massive black holes are
actively accreting Seyferts than those of lower mass.

\begin{figure*}
\begin{center}
\leavevmode
\psfig{figure=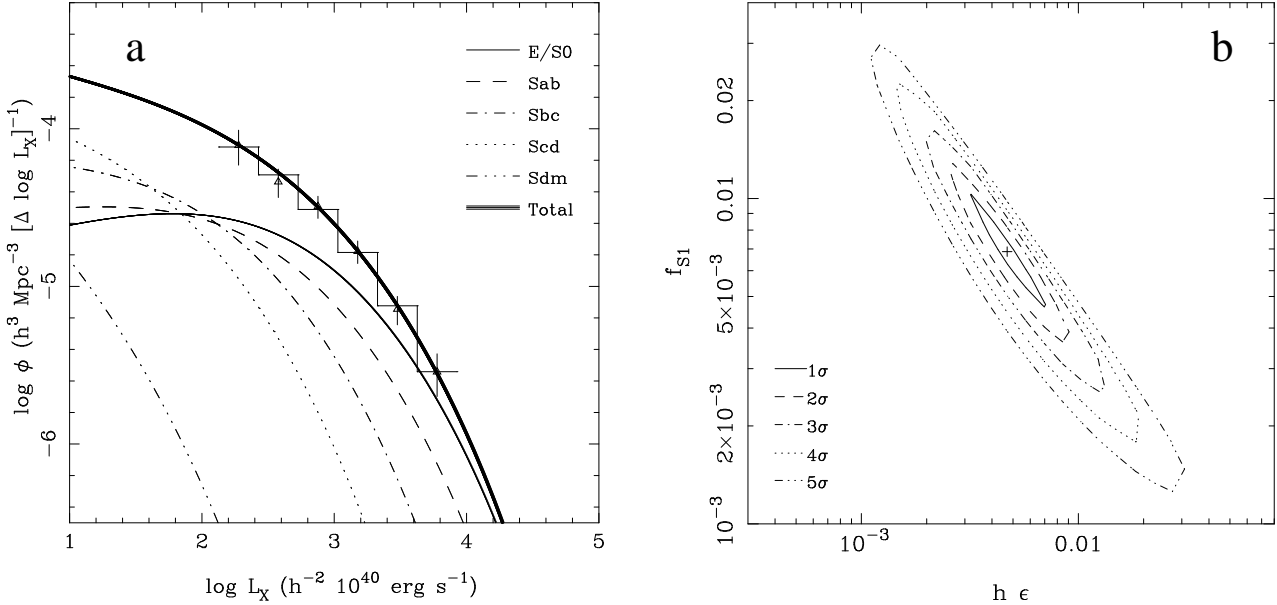,height=80truemm,angle=270}
\caption{(a) X--ray luminosity function of the EMSS $z<0.2$ AGN and best fit 
model in which the Seyfert 1 fraction of E and S0 galaxies is only half that in
 other types, and the Seyfert 1 nuclei hosted by E and S0 galaxies accrete with
only half the Eddington ratio of Seyfert nuclei in later galaxy types 
(Model B).
(b) $\chi^{2}$ confidence contours for the values of Eddington ratio
$\epsilon$ and Seyfert 1 fraction $f_{S1}$ for Sab and later type galaxies in Model B.}
\label{fig:adhoclumifun}
\end{center}
\end{figure*}

An ad-hoc example of one of these models is shown in
Figs. \ref{fig:adhoclumifun}a. In this model (hereafter Model B) we
have assumed that the fraction of active Seyferts in E/S0 galaxies,
and their Eddington ratios, are only half that of later type
galaxies. Again, the best fit luminosity function is an extremely good
fit with a \(\chi^{2}\) of 0.45, but this time the AGN luminosity
function is produced by AGN with a more plausible mix of host galaxy
morphologies. The fitted parameters are \(f_{S1} = 7\pm3 \times
10^{-3}\) and \(h \epsilon = 5\pm2 \times 10^{-3}\) for Sab and later
galaxies (and by design half these values for E and S0 galaxies).

\begin{figure*}
\begin{center}
\leavevmode
\psfig{figure=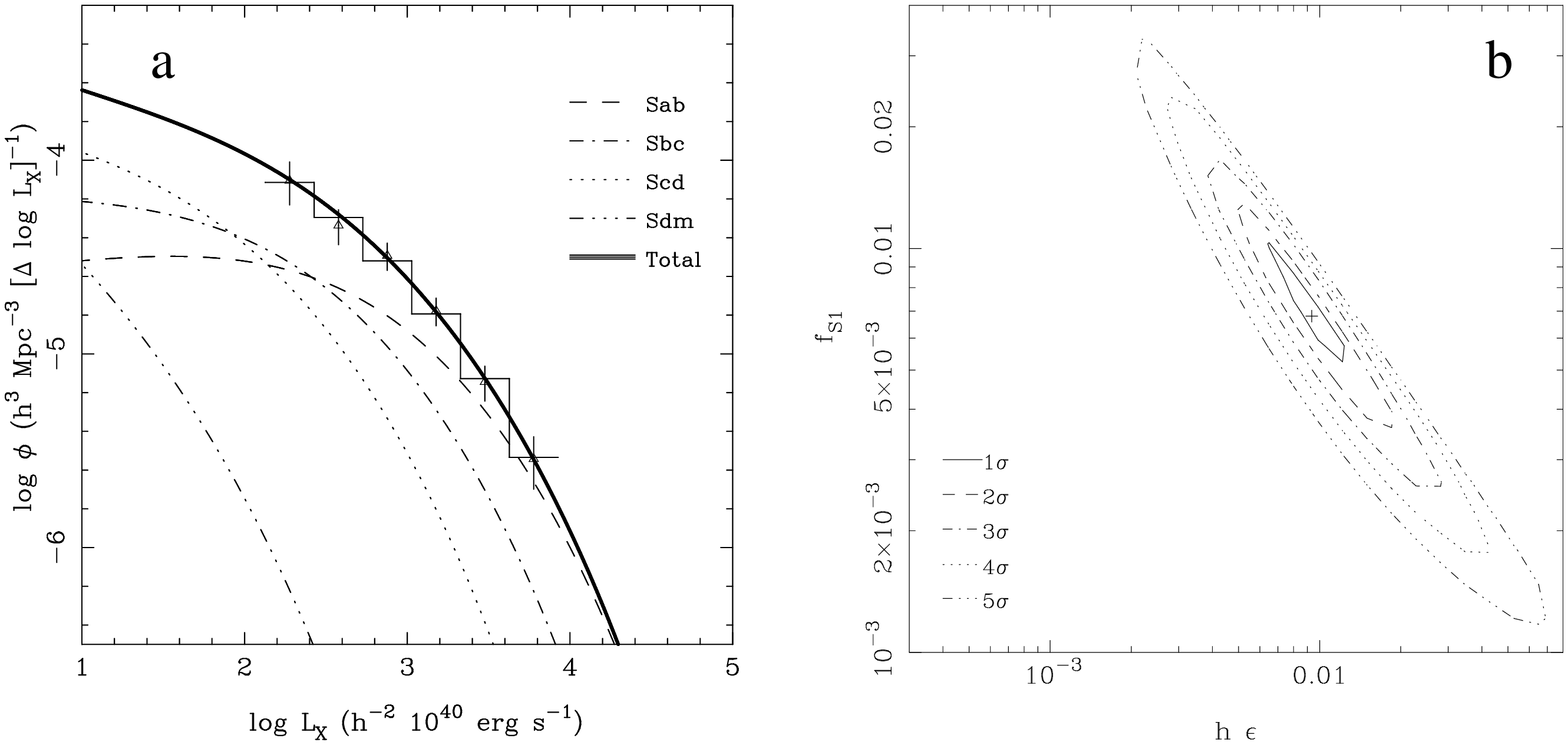,height=80truemm,angle=270}
\caption{(a) X--ray luminosity function of the EMSS $z<0.2$ AGN and
best fit model in which E and S0 galaxies host no AGN (Model C).  (b)
$\chi^{2}$ confidence contours for the values of Eddington ratio
$\epsilon$ and Seyfert 1 fraction $f_{S1}$ for Sab and later type
galaxies in Model C.}
\label{fig:modelclumifun}
\end{center}
\end{figure*}

To obtain a conservative limit on the range of parameter space that
brackets \(f_{S1}\) and \(\epsilon\), we have also considered the
opposite extreme to the E/S0 dominated Model A: in Model C the active
fraction of E and S0 galaxies is set to zero and they therefore make no contribution whatsoever to the AGN luminosity function.  Since the fraction of Seyfert nuclei which are hosted by E and S0 galaxies is certainly between 0 and 1, it is reasonable to expect that the real values of \(f_{S1}\) and \(\epsilon\) must lie somewhere between the acceptable values for Model A and Model C. The best fit AGN luminosity function and \(\Delta \chi^{2}\) confidence contours for Model C are shown in Fig. \ref{fig:modelclumifun}.

\section{Discussion}

The approach used here easily reproduces the shape of the AGN
luminosity function; models A, B and C all have extremely low
\(\chi^{2}\).  The steepening of the AGN luminosity function from low
to high luminosity is inherited from the black hole mass function,
which in turn inherits the shape from the galaxy luminosity
functions. Hence the shape of the AGN luminosity function ultimately
derives from the same physical processes that give rise to the shapes
of galaxy luminosity functions.

The fitted Seyfert 1 fraction in Sab and later type galaxies for Model
B is \(\sim 0.5 - 1 \%\). This is quite a robust result: it isn't
strongly affected by the contribution of E and S0 galaxies because
Models A and C have very similar best fit values for \(f_{S1}\).  This
compares well with the findings of optical emission line surveys of
local galaxies.  Maiolino \& Rieke (1995), find that 5\% of the
revised Shapley-Ames catalogue of galaxies are Seyferts (with a
Seyfert 1: Seyfert 2 ratio of 1:4 this corresponds to a Seyfert 1
fraction of 1\%).  Ho, Filippenko \& Sargent (1997), find a Seyfert
fraction which is twice that found by Maiolino \& Rieke (1995), and
therefore (assuming the Maiolino \& Rieke Seyfert 1: Seyfert 2 ratio)
about a factor two higher than would be expected from our value of
\(f_{S1}\).  However, their Seyfert detection rate is high because
their survey is sensitive to objects with very weak emission lines,
which they term ``dwarf'' Seyferts; it is not yet known whether these
objects have the same X--ray properties as more luminous Seyferts
which make up the EMSS sample.
Both Maiolino \& Rieke (1995) and Ho, Filippenko \& Sargent (1997)
find that the most common morphology for Seyfert galaxies is Sa - Sb
further justifying our rejection of model A in Section
\ref{sec:fitting}.

In contrast to \(f_{S1}\), the value of \(\epsilon\) for typical Sab
Seyfert galaxies depends quite strongly on the Seyfert activity in E
and S0 galaxies. Taking the extremes of models A and C, a conservative
conclusion is that \(0.001 < \epsilon < 0.02\) for Seyfert nuclei in
Sab galaxies. Our reasonable model B suggests \(\epsilon \sim
0.005\) should be typical, considerably sub-Eddington and consistent
with Seyfert activity occurring recurrently in a significant fraction
of galaxies (model `R' from Cavaliere \& Padovani 1988). 
This is in good agreement with the predictions of recent
models in which AGN, with lifetimes of order \(10^{7}\) years, are
produced during the formation and merging of galaxies in a cold dark
matter dominated Universe: Kauffmann \& Haehnelt (2000) predict
typical \(\epsilon \sim 0.01\) while Haiman \& Menou (2000) predict
typical \(\epsilon \sim 0.001\).  Our results are also consistent with
the range of Seyfert Eddington ratios given in the recent compilation
by Wandel (1999). These lie between 0.001 and 1 and were obtained
using a variety of different methods: broad line region kinematics,
X--ray variability and modelling accretion disk spectra.

However, the range of acceptable
\(\epsilon\) is too low to be consistent with accretion disk models
which have large outbursts.  Siemiginowska \& Elvis (1997) show that
if AGN accretion disks are subject to the thermal-viscous instability
driven by hydrogen ionization (Lin \& Shields 1986) they are probably
only observed in outburst, with typical \(\epsilon \sim 0.1\). The
study by Burderi, King \& Szuszkiewicz (1998) concludes that if AGN
have optically thick, geometrically thin accretion disks (Shakura \&
Sunyaev 1973) they almost certainly are subject to this kind of
instability. The results presented here would therefore suggest that
AGN are not fuelled by standard thin disks.

As explained in Section \ref{sec:fitting}, the only way to get a
plausible distribution of Seyfert host galaxies is to model the more
massive black holes in early type galaxies with smaller values of
\(\epsilon\), \(f_{S1}\) or both. Both these alternatives have
important consequences for our understanding of AGN behaviour. 

The first (\(\epsilon \) smaller for more massive AGN) is intuitively
reasonable because the early type galaxies which host more massive AGN
contain less gas with which to feed them. However, it is contrary to
the results of Wandel \& Petrosian (1988) and Sun \& Malkan (1989),
who find from accretion disk modelling that luminous QSOs are both
more massive, and accrete with higher Eddington ratios, than Seyfert
galaxies (although this could be interpreted as a dependence of
Eddington ratio on redshift rather than mass).  The AGN evolution
models of Kauffmann \& Haehnelt (2000) and Salucci \etal (1999), also
favour a situation in which the more luminous present epoch AGN are more
massive {\it and} have higher Eddington ratios.

The second way to obtain reasonable Seyfert morphologies, by modelling
a smaller Seyfert fraction \(f_{S1}\) for more massive black holes, is
less controversial. For example, both Maiolino \& Rieke (1995) and Ho,
Filippenko \& Sargent (1997) find that a higher fraction of Sa-Sb
galaxies have Seyfert nuclei than E/S0 galaxies. In this case the mass
function of active nuclei must be steeper than the mass function of
inactive nuclei, unlike the simple recurrent activity models 
represented by model `R' in figure 1 of Cavaliere \& Padovani (1988). 

This investigation can be substantially improved upon with a sample of
X--ray selected Seyferts with optical morphologies; this would allow
the matching of galaxy and AGN luminosity functions individually for
each morphological type, providing a much more rigorous solution than
our model B.  This obviously requires a larger AGN sample than the
EMSS one used here, for which the luminosity functions of different
morphological types would be dominated by Poisson noise. We can expect
that such a sample may soon be available from \ros\ All Sky Survey
optical identification programmes.

\section{Acknowledgments}

We would like to thank Jason Stevens, Francisco Carrera, Gavin Ramsay, 
Roberto Soria and Keith Mason for useful comments on the draft version of 
this paper.

\section{References}

\refer Boyle B.J., Shanks T., \& Peterson B.A., 1988, MNRAS, 235, 935

\refer Burderi L., King A. R., Szuszkiewicz E., 1998, ApJ, 509, 85

\refer Cattaneo A., Haehnelt M.G., Rees M.J., 1999, MNRAS, 308, 77

\refer Cavaliere A., \& Padovani P., 1988, ApJ, 333, L33

\refer de Vaucouleurs G., 1959, in {\it Handbuch der Physik}, vol 53, ed. Fl\"
ugge S., (Berlin: Springer) p.275
 
\refer Dunlop J.S., \& Peacock J.A., 1990, MNRAS, 247, 19

\refer Elvis M., \etal, 1994, ApJS, 95, 1

\refer Fabian A.C., 1999, MNRAS, 308, L39

\refer Folkes S., \etal, 1999, MNRAS, 308, 459

\refer Haehnelt M.G., \& Rees M.J., MNRAS, 1993, 263, 168

\refer Haiman Z., \& Menou K., 2000, ApJ, 531, 42

\refer Ho L., Filippenko A.V., \& Sargent W.L.W., 1997, ApJ, 487, 568

\refer Kauffmann G., \& Haehnelt M., 2000, MNRAS, 311, 576

\refer Kormendy J., \& Richstone D., 1995, Annu. Rev. Astron. Astrophys., 33,
581

\refer Lynden-Bell D., 1969, Nat, 223, 690

\refer Magorrian J., \etal 1998, ApJ, 115, 2285

\refer Meisels A., \& Ostriker J.P., 1984, AJ, 89, 1451

\refer Maiolino R., \& Rieke G.H., 1995, ApJ, 454, 95

\refer Monaco P., Salucci P., \& Danese L., MNRAS, 311, 279

\refer Page M.J., Mason K.O., McHardy I.M., Jones L.R., Carrera F.J., 1997,
MNRAS, 291, 324

\refer Page M.J., \& Carrera F.J., 2000, MNRAS, 311, 433

\refer Pence W., 1976, ApJ, 203, 39

\refer Rees M.J., 1984, Annu. Rev. Astron. Astrophys., 22, 471

\refer Salucci P., Szuszkiewicz E., Monaco P., Danese L., 1999, MNRAS, 307, 637

\refer Shakura N.I., \& Sunyaev R.A., 1973, A\&A, 24, 337

\refer Siemiginowska A., \& Elvis M., 1997, ApJ, 482, L9

\refer Stocke J.T., \etal, 1991, ApJS, 76, 813

\refer Sun W.H., \& Malkan M.A., 1989, ApJ, 346, 68

\refer Wandel A., \& Petrosian V., 1988, ApJ, 329, L11

\refer Wandel A., 1999, in ``Structure and Kinematics of Quasar Broad
Line Regions'', ASP Conference Series, Vol. 175, Ed. C. M. Gaskell,
W. N. Brandt, M. Dietrich, D. Dultzin-Hacyan, and M. Eracleous, p.213

\end{document}